\documentclass[english]{article}
\usepackage[T1]{fontenc}
\usepackage[latin9]{inputenc}
\usepackage[letterpaper]{geometry}
\usepackage{babel}
\usepackage{units}
\usepackage{mathrsfs}
\usepackage{bm}
\usepackage{amsmath}
\usepackage{amssymb}
\usepackage{esint}
\usepackage[unicode=true]
 {hyperref}

\makeatletter
\usepackage{cite}
\usepackage{tensind}
\tensordelimiter{?}
\usepackage{bm}
\usepackage{tocloft} 

\makeatother

\begin{document}

\title{General Relativity Revisited: Generalized Nordström Theory}

\author{Johan Bengtsson\\
\\
Member of Center for Accelerator Science and Education\\
(Stony Brook University and Brookhaven National Laboratory)}

\date{09/14/16}

\maketitle
\begin{center}
\vspace{1in}

\par\end{center}
\begin{abstract}
\noindent In 1945 Einstein concluded that \cite{Meaning_of_Rel.}:
\textquotedblleft{}\textit{The present theory of relativity is based
on a division of physical reality into a metric field (gravitation)
on the one hand, and into an electromagnetic field and matter on the
other hand. In reality space will probably be of a uniform character
and the present theory be valid only as a limiting case. For large
densities of field and of matter, the field equations and even the
field variables which enter into them will have no real significance.}\textquotedblright{}.
The dichotomy can be resolved by introducing a scalar field/potential
algebraically related to the Ricci tensor for which the corresponding
metric is free of additional singularities. Hence, although a fundamentally
nonlinear theory, the scalar field/potential provides an analytic
framework for interacting particles; described by linear superposition.
The stress tensor for the scalar field includes both the sources of
and the energy-momentum for the gravitational field, and has zero
covariant and ordinary divergence. Hence, the energy-momentum for
the gravitational field and sources are conserved. The theory's predictions
agree with the experimental results for General Relativity. By introducing
the corresponding Lagrangian to analytic mechanics, what is experimentally
known for GR can be accounted for.\addcontentsline{toc}{section}{\textbf{Abstract}}

\newpage{}

\tableofcontents{}\newpage{}
\end{abstract}

\section{Introduction: Towards a Theory of General Relativity}

\subsection{Outline}

After A. Einstein introduced Special Relativity (SR) 1905\cite{Special_Rel.},
the general case, i.e., a General Theory of Relativity (GR) turned
out to be elusive. Eventually, he made progress through a collaboration
with M. Grossmann 1913 \cite{Entwurf} who introduced him to the tensor
calculus developed by: K. Gauss 1827 \cite{Gauss}, B. Riemann 1854
\cite{Riemann}, E. Christoffel 1865 \cite{Christoffel}, and G. Ricci
and T. Levi-Civita 1901 \cite{Levi-Civita-Ricci}. According to Einstein,
his student friend was somewhat reluctant (p. 152 \cite{Einstein-Grossmann,Weinstein-1}):
\begin{quote}
``\textit{...with the restriction that he would not be responsible
for any statements and won't assume any interpretations of physical
nature.}''
\end{quote}
Similarly, in a letter to Levi-Civita 1917, regarding a controversy
on the gravitational stress tensor \cite{Levi-Civita_Stress_Tensor},
Einstein wrote \cite{Italians}:
\begin{quote}
``\textit{I admire the elegance of your method and calculation. It
must be nice to ride through these fields upon the horse of true mathematics,
while the likes of us have to make our way laboriously on foot...}''
\end{quote}
At the same time D. Hilbert was developing a general mathematical
framework for field theories \cite{Hilbert}. So, shortly after pusblishing
his 1916 GR paper \cite{Einstein_GR}, Einstein re-derived his field
equations using Hilbert's variational approach \cite{Einstein_Ham}.
Given the time it took to establish a General Theory of Relativity,
it is no surprise that the path was paved by a whole slew of intellectual
explorers at a time it was ripe; e.g. space-time continuum, etc. For
example, G. Nordström was one of the first to contribute with a self-consistent
approach \cite{Nordstr=0000F6m,Pais}.

The outline of this essay is as follows:
\begin{enumerate}
\item The variational framework is used to generate:\\
\\
a) Einstein's field equations,\\
b) equations of motion for a perfect fluid,\\
c) field equations for a massless scalar field.
\item An (differential) algebraic relation between the scalar field stress
tensor and the Ricci tensor is obtained.
\item The effect on the metric is given by the Christoffel symbols. The
metric is obtained by solving the system for an $\nicefrac{1}{r}$
potential.
\item By evaluating and expanding the corresponding Lagrangian for analytic
mechanics, it is transparent that the theory's predictions agree with
GR; to first order in the gravitational constant.
\item The theory is essentially a generalization of Nordström's theory from
1913 \cite{Nordstr=0000F6m,Pais,Einstein_Nordstr=0000F6m}; i.e.,
a scalar field in curved space-time vs. Minkowski metric.
\end{enumerate}

\subsection{Hilbert's Variational Framework}

The action $S$ for a Lagrangian density $\mathscr{L}$ is \cite{Hilbert}

\begin{equation}
S=\int\mathscr{L}d^{4}x.
\end{equation}
Variation of the metric $g_{ij}\rightarrow g_{ij}+\delta g_{ij}$
generates the canonical stress tensor density

\begin{equation}
\mathscr{T}^{ij}\equiv\sqrt{-g}T^{ij}=-2\frac{\delta\mathscr{L}}{\delta g_{ij}}=-2\sqrt{-g}\frac{\delta\left(\frac{1}{\sqrt{-g}}\mathscr{L}\right)}{\delta g_{ij}}+g^{ij}\mathscr{L}
\end{equation}
which has zero covariant divergence
\begin{equation}
\nabla_{r}\mathscr{T}^{ri}=\sqrt{-g}\nabla_{r}T^{ri}=0
\end{equation}
where $\nabla_{i}$ is the covariant derivative, since (e.g. \cite{Landau-Lifshitz})
\begin{equation}
\int\frac{\delta\mathscr{L}}{\delta g_{rs}}\delta g^{rs}d^{4}x=\int\sqrt{-g}\left(\nabla_{r}?T^{r}{}_{s}?\right)\delta x^{s}d^{4}x=0.
\end{equation}
However, the integral

\begin{equation}
\int\mathscr{T}^{k}dS_{k}
\end{equation}
is conserved only if the ordinary divergence is zero

\begin{equation}
\partial_{r}\mathscr{T}^{ri}=0.
\end{equation}

\subsection{General Relativity}

For General Relativity (GR) \cite{Einstein_Ham}

\begin{equation}
\mathscr{L}=\frac{1}{2\kappa}\sqrt{-g}R+\mathscr{L}_{\mathrm{m}},\qquad\kappa\equiv\frac{8\pi G}{c_{0}^{4}}
\end{equation}
where $\mathscr{L}_{\mathrm{m}}$ is the Lagrangian density describing
matter. A variation of the metric gives
\begin{equation}
\delta S=\delta\int\mathscr{L}d^{4}x=\delta\int\left(\frac{1}{2\kappa}\sqrt{-g}g^{ir}R_{rj}+\mathscr{L}_{\mathrm{m}}\right)d^{4}x=\frac{\sqrt{-g}}{2\kappa}\left(R_{rs}-\frac{1}{2}g_{rs}R-\kappa T_{rs}\right)g^{rs}d^{4}x=0
\end{equation}
which leads to Einstein's field equations \cite{Einstein_GR}

\begin{equation}
R_{ij}-\frac{1}{2}g_{ij}R=\kappa T_{ij}^{\mathrm{\left(m\right)}}\label{eq: Einstein's Field Eqs.}
\end{equation}
with the contraction

\begin{equation}
R=-\kappa T^{\mathrm{\left(m\right)}}.\label{eq: Einstein's Contr. Field Eqs.}
\end{equation}
For a static spherically symmetric body with mass $M$ they have the
solution (Schwarzschild metric) \cite{Schwarzschild}
\begin{equation}
ds^{2}=-\left(1+\frac{2\phi}{c_{0}^{2}}\right)^{2}c_{0}^{2}dt^{2}+\frac{dr^{2}}{\left(1-\frac{2\phi}{c_{0}^{2}}\right)^{2}}+r^{2}d\theta^{2}+r^{2}\sin^{2}\left(\theta\right)d\varphi^{2}\label{eq: Schwarzschild Metric}
\end{equation}
where
\begin{equation}
\phi=-\frac{GM}{r},\qquad r>0.
\end{equation}
It becomes singular for (Schwarzschild radius)
\begin{equation}
r=\frac{2GM}{c_{0}^{2}}.
\end{equation}

\subsection{Gravitational Stress Tensor: The Einstein - Levi-Civita Controversy}

Levi-Civita, inspired by Einstein's 1916 theory \cite{Einstein_GR},
the same year introduced the notion parallel transport \cite{Levi-Civita_Parallel_Transp}
to simplify the geometric framework.

The Einstein - Levi-Civita controversy 1917 \cite{Levi-Civita_Stress_Tensor}
was regarding the interpretation of the Einstein equations
\begin{equation}
R_{ij}-\frac{1}{2}g_{ij}R=\kappa T_{ij}^{\left(\mathrm{m}\right)}.
\end{equation}
In particular, Einstein viewed them as field equations whereas Levi-Civita
wrote them in the form
\begin{equation}
T_{ij}^{\left(\mathrm{g}\right)}+T_{ij}^{\left(\mathrm{m}\right)}=0
\end{equation}
where
\begin{equation}
T_{ij}^{\left(\mathrm{g}\right)}=\frac{1}{\kappa}\left(R_{ij}-\frac{1}{2}g_{ij}R\right)\label{eq: Levi-Civita}
\end{equation}
and interpreted $T_{ij}^{\left(\mathrm{g}\right)}$ as the stress
tensor, originating exclusively from the metric $ds^{2}$, that balances
all mechanical actions. Instead, Einstein, introduced an ad-hoc (non-symmetric)
pseudo-tensor for the gravitational field Eq. (50)\cite{Einstein_GR}
(see also section 87 in ref. \cite{Tolman})
\begin{align}
?t^{i}{}_{j}?^{\left(\mathrm{g}\right)} & \equiv\frac{1}{2\kappa}\left[G^{\star}\delta_{j}^{i}-\left(\partial_{\partial_{j}g^{rs}}G^{\star}\right)\partial_{i}g^{rs}\right]\nonumber \\
 & =\frac{1}{2\kappa}\delta_{j}^{i}g^{rs}?\Gamma^{t}{}_{ru}??\Gamma^{u}{}_{st}?-g^{rs}?\Gamma^{i}{}_{ru}??\Gamma^{u}{}_{sj}?
\end{align}
where (affine scalar; only invariant under linear transformations)
\begin{equation}
G^{\star}=g^{rs}\left(?\Gamma^{t}{}_{ru}??\Gamma^{u}{}_{st}?-?\Gamma^{t}{}_{rs}??\Gamma^{u}{}_{tu}?\right)
\end{equation}
and it follows that
\begin{equation}
\partial_{r}\left(\sqrt{-g}?T^{r}{}_{i}?^{\left(\mathrm{m}\right)}+?t^{r}{}_{i}?^{\left(\mathrm{g}\right)}\right)=0.
\end{equation}
In particular
\begin{equation}
\int\sqrt{-g}Rd^{4}x=\int\sqrt{-g}G^{\star}d^{4}x+\mathrm{surface\: term}
\end{equation}
and $G^{\star}$ depends only on $g_{ij}$ and their first derivatives;
since $R$ is linear in the second derivatives.

Similarly, in 1947 Landau-Lifshitz introduced the (symmetric) pseudo-tensor
\cite{Landau-Lifshitz}
\begin{align*}
t^{ij\left(\mathrm{g}\right)}= & \frac{1}{2\kappa}\left[\left(2?\Gamma^{t}{}_{rs}??\Gamma^{u}{}_{tu}?+?\Gamma^{t}{}_{ru}??\Gamma^{u}{}_{st}?-?\Gamma^{t}{}_{rt}??\Gamma^{u}{}_{su}?\right)\left(g^{ir}g^{js}-g^{ij}g^{rs}\right)\right.\\
 & +g^{ir}g^{st}\left(?\Gamma^{j}{}_{ru}??\Gamma^{u}{}_{st}?+?\Gamma^{j}{}_{st}??\Gamma^{u}{}_{ru}?-?\Gamma^{j}{}_{tu}??\Gamma^{u}{}_{rs}?-?\Gamma^{j}{}_{rs}??\Gamma^{u}{}_{tu}?\right)\\
 & +g^{jr}g^{st}\left(?\Gamma^{i}{}_{ru}??\Gamma^{u}{}_{st}?+?\Gamma^{i}{}_{st}??\Gamma^{u}{}_{ru}?-?\Gamma^{i}{}_{rt}??\Gamma^{u}{}_{su}?-?\Gamma^{i}{}_{rt}??\Gamma^{u}{}_{su}?\right)\\
 & +g^{rs}g^{tu}\left(?\Gamma^{i}{}_{rt}??\Gamma^{j}{}_{su}?-?\Gamma^{i}{}_{rs}??\Gamma^{j}{}_{tu}?\right)
\end{align*}
for which
\begin{equation}
\partial_{r}\left(-g\left(T^{ir\left(\mathrm{m}\right)}+t^{ir\left(\mathrm{g}\right)}\right)\right)=0.
\end{equation}

\subsection{Perfect Relativistic Fluid}

A Lagrangian approach for a perfect relativistic fluid requires variational
calculus with constraints \cite{Var_Calc}%
\footnote{Similarly, Maxwell's equations requires the constraint $\delta\nabla_{r}F^{ri}=0$;
or the introduction of a vector potential.%
}. For example, starting from\cite{Dirac}

\begin{equation}
\mathscr{L}_{\mathrm{m}}\left(p^{i}\right)=-c_{0}\sqrt{-g}\rho\sqrt{-u_{r}u^{r}}=-c_{0}\sqrt{-p_{r}p^{r}},\qquad p^{i}\equiv\sqrt{-g}\rho u^{i}
\end{equation}
and (mass conservation)
\begin{align*}
\partial_{r}p^{r} & =\partial_{r}\left(\sqrt{-g}\rho u^{r}\right)=\sqrt{-g}\nabla_{r}\left(\rho u^{r}\right)=\nabla_{r}p^{r}=0\Rightarrow\\
\frac{d\rho}{ds} & =u^{r}\partial_{r}\rho=u^{r}\nabla_{r}\rho=-\rho\nabla_{r}u^{r}.
\end{align*}
variation of the momenta $p^{i}\rightarrow p^{i}+\delta p^{i}$ and
using the relations
\begin{align}
\delta\sqrt{-p_{r}p^{r}} & =-\frac{1}{\sqrt{-p_{r}p^{r}}}p_{r}\delta p^{r}=\frac{1}{c_{0}}u_{r}\delta p^{r},\nonumber \\
\delta p^{i} & =\partial_{r}\left(p^{r}\delta x^{i}-p^{i}\delta x^{r}\right),\nonumber \\
\partial_{i}u_{j}-\partial_{j}u_{i} & =\nabla_{i}u_{j}-\nabla_{j}u_{i},\\
\nabla_{i}\left(u^{r}u_{r}\right) & =2u^{r}\nabla_{i}u_{r}=0
\end{align}
leads to

\begin{equation}
\delta S=\delta\int\mathscr{\mathscr{L}_{\mathrm{m}}}d^{4}x=\int\sqrt{-g}\rho u^{r}\left(\nabla_{r}u_{s}\right)\delta x^{s}d^{4}x
\end{equation}
and the equations of motion (geodesic)

\begin{equation}
u^{r}\nabla_{r}u^{i}=\frac{du^{i}}{ds}+?\Gamma^{i}{}{}_{rs}?u^{r}u^{s}=0.\label{eg: Eqs. of Motion Fluid}
\end{equation}
The canonical stress tensor density (for $p^{i}$) is

\begin{equation}
\mathscr{T}^{ij\left(m\right)}=-2\frac{\delta\mathscr{L}\left(p^{i}\right)}{\delta g_{ij}}=2c_{0}\frac{\delta\sqrt{-p_{r}p^{r}}}{\delta g_{ij}}=-\sqrt{-g}\rho u^{i}u^{j}
\end{equation}
i.e., Eq (25.4) in ref. \cite{Dirac}, with the trace

\begin{equation}
?{\mathscr{T}}{}^{r}{}{}_{r}?^{\left(m\right)}=\sqrt{-g}\rho c_{0}^{2}\label{eq: Trace Fluid}
\end{equation}
and the covariant divergence is zero (equations of motion)

\begin{equation}
\nabla_{r}?{\mathscr{T}}^{r}{}{}_{i}?^{\left(m\right)}=-\partial_{r}\left(\sqrt{-g}\rho u^{r}u_{i}\right)=-\nabla_{r}\left(p^{r}u_{i}\right)=u_{i}\nabla_{r}p^{r}+\sqrt{-g}\rho u^{r}\nabla_{r}u_{i}=0
\end{equation}
i.e., for mass conservation $\nabla_{r}p^{r}=0$ and along a geodesic
$u^{r}\nabla_{r}u_{i}=0$.

Alternatively, one may start from the Lagrangian density
\begin{equation}
\mathscr{L}_{\mathrm{m}}\left(u^{i}\right)=-\frac{1}{2}p^{r}u_{r},
\end{equation}
since
\begin{equation}
\mathscr{L}_{p}\equiv-\int p^{r}du_{r}=-\int\sqrt{-g}\rho u^{r}du_{r}=-\frac{1}{2}p^{r}u_{r},
\end{equation}
to generate the canonical stress tensor
\begin{equation}
?{\mathscr{T}}^{i}{}_{j}?^{\left(m\right)}=-2\sqrt{-g}\frac{\delta\left(\frac{1}{\sqrt{-g}}\mathscr{L}\right)}{\delta g_{ij}}+g^{ij}\mathscr{L}=\sqrt{-g}\rho\left(u^{i}u_{j}+\frac{c_{0}^{2}}{2}\delta_{j}^{i}\right)\label{eq: Stress Tensor Fluid}
\end{equation}
and proceed directly to the equations of motion (see last paragraph
of section 4.c in ref. \cite{Var_Calc})
\begin{equation}
\nabla_{r}?{\mathscr{T}}^{r}{}_{i}?^{\left(m\right)}=\nabla_{r}\left(\sqrt{-g}\rho u^{r}u_{i}+\frac{c_{0}^{2}}{2}\sqrt{-g}\rho\delta_{i}^{r}\right)=u_{i}\nabla_{r}p^{r}+\sqrt{-g}\rho u^{r}\nabla_{r}u_{i}+\frac{c_{0}^{2}\sqrt{-g}}{2}\nabla_{i}\rho=0.
\end{equation}
For a fluid for which (mass conservation)
\begin{equation}
\nabla_{r}p^{r}=0
\end{equation}
they simplify to
\begin{equation}
u^{r}\nabla_{r}u^{i}=\frac{du^{i}}{ds}+?\Gamma^{i}{}{}_{rs}?u^{r}u^{s}+\frac{c_{0}^{2}}{2}g^{ir}\partial_{r}\ln\left(\rho\right)=0\label{eq: Eqs. of Motion Fluid 2}
\end{equation}
which differ from Eq. (\ref{eg: Eqs. of Motion Fluid}) by the last
term.

For an incompressible fluid $\rho=\mathrm{const.}$
\begin{equation}
u^{r}\nabla_{r}u^{i}=\frac{du^{i}}{ds}+?\Gamma^{i}{}{}_{rs}?u^{r}u^{s}=0
\end{equation}

\section{Massless Scalar Field}

\subsection{Lagrangian Density and Field Equations}

For a free, massless scalar field

\begin{equation}
\mathscr{L}\left(\phi,\phi_{i}\right)=\frac{1}{8\pi G}\sqrt{-g}\phi^{r}\phi_{r}=\frac{1}{8\pi G}\sqrt{-g}g^{rs}\phi_{s}\phi_{r},\qquad\phi_{i}\equiv\partial_{i}\phi.
\end{equation}
A (covariant) variation of the field $\phi\rightarrow\phi+\delta\phi$
gives (Euler-Lagrange equations) \cite{Tupper}

\begin{equation}
\partial_{\phi}\mathscr{L}-\partial_{r}\left(\partial_{\phi_{r}}\mathscr{L}\right)=0\label{eq: Scalar Field Eqs.}
\end{equation}
which leads to the field equations
\begin{equation}
\partial_{r}\left(\sqrt{-g}\phi^{r}\right)=\sqrt{-g}\nabla_{r}\phi^{r}=\frac{1}{2}\partial_{\phi}\left(\sqrt{-g}\phi^{r}\phi_{r}\right)=\frac{1}{2}\left(\partial_{\phi}\sqrt{-g}\right)\phi^{r}\phi_{r}
\end{equation}
since $\partial_{\phi}\left(\phi^{r}\phi_{r}\right)=0$. The stress
tensor density is

\begin{equation}
\mathscr{T}^{ij\left(\phi\right)}=-2\frac{\delta\mathscr{L}}{\delta g_{ij}}=-\frac{\sqrt{-g}}{4\pi G}\left(\phi^{i}\phi^{j}-\frac{1}{2}g^{ij}\phi^{r}\phi_{r}\right)=-\phi_{j}\partial_{\phi_{i}}\mathscr{L}+g^{ij}\mathscr{L}.\label{eq: Stress Tensor Scl Field}
\end{equation}
with the trace

\begin{equation}
?{\mathscr{T}}^{r}{}{}_{r}?^{\left(\phi\right)}=\frac{\sqrt{-g}}{4\pi G}\phi^{r}\phi_{r}\label{eq: Trace Scl Field}
\end{equation}
and zero covariant as well as ordinary divergence
\begin{equation}
\partial_{r}?{\mathscr{T}}^{r}{}{}_{i}?^{\left(\phi\right)}=\partial_{r}\left(-\phi_{j}\partial_{\phi_{i}}\mathscr{L}+\delta_{j}^{i}\mathscr{L}\right)=0\label{eq: Zero Trace}
\end{equation}
where we have used the Euler-Lagrange equations Eq. (\ref{eq: Scalar Field Eqs.})
and

\begin{align}
\delta\mathscr{L} & =\partial_{\phi}\mathscr{L}\delta\phi+\partial_{\phi_{r}}\mathscr{L}\delta\phi_{r}\Rightarrow\\
\partial_{i}\mathscr{L} & =\left(\partial_{\phi}\mathscr{L}\right)\phi_{i}+\left(\partial_{\phi_{r}}\mathscr{L}\right)\partial_{i}\phi_{r}=\left(\partial_{\phi}\mathscr{L}\right)\phi_{i}+\left(\partial_{\phi_{r}}\mathscr{L}\right)\partial_{r}\phi_{i}.
\end{align}

By using the expression for the trace, the field equations can be
written
\begin{equation}
\nabla_{r}\left(\sqrt{-g}\phi^{r}\right)=\frac{1}{2}\left(\partial_{\phi}\sqrt{-g}\right)\phi^{r}\phi_{r}=\frac{2\pi G}{\sqrt{-g}}\left(\partial_{\phi}\sqrt{-g}\right)?{\mathscr{T}}^{r}{}{}_{r}?^{\left(\phi\right)}=2\pi G\left(\partial_{\phi}\ln\left(\sqrt{-g}\right)\right)?{\mathscr{T}}^{r}{}{}_{r}?^{\left(\phi\right)}
\end{equation}
i.e., the RHS is the trace of the stress tensor density.

By introducing the conformal factor
\begin{equation}
\sqrt{-g}=e^{-\nicefrac{2\phi}{c_{0}^{2}}}\label{eq: Conformal Factor}
\end{equation}
they simplify to (covariant d'Alambert equation)
\begin{equation}
\nabla_{r}\left(\sqrt{-g}\phi^{r}\right)=\nabla_{r}\left(\bm{{\phi}}^{r}\right)=\square\left(\bm{{\phi}}\right)=-\frac{4\pi G}{c_{0}^{2}}?{\mathscr{T}}^{r}{}{}_{r}?^{\left(\phi\right)}=-\frac{4\pi G}{c_{0}^{2}}\sqrt{-g}?T^{r}{}{}_{r}?^{\left(\phi\right)}\label{eq: Covariant d'Alambert Eq.}
\end{equation}
where we have introduced the scalar field density $\bm{{\phi}}^{i}$;
the dual field to $\phi_{i}$. Hence, the metric $g_{ij}$ is a constitutive
relation describing the properties of the space-time continuum
\begin{equation}
\bm{{\phi}}^{i}=\sqrt{-g}g^{ir}\phi_{r}
\end{equation}
between the abstract mathematical field $\phi_{i}$ and measurable
physical field $\bm{{\phi}}^{i}$.

By inserting the stress tensor Eq. (\ref{eq: Stress Tensor Scl Field})
into the R.H.S. of Einstein's field Eqs. (\ref{eq: Einstein's Field Eqs.})

\begin{equation}
?R^{i}{}_{j}?-\frac{1}{2}\delta_{j}^{i}R=\kappa?T^{i}{}_{j}?=-\frac{2}{c_{0}^{4}}\left(\phi^{i}\phi_{j}-\frac{1}{2}\delta_{j}^{i}\phi^{r}\phi_{r}\right)\label{eq: Scalar - GR}
\end{equation}
one obtains the algebraic relation \cite{Yilmaz}

\begin{equation}
?R^{i}{}_{j}?=-\frac{2}{c_{0}^{4}}\phi^{i}\phi_{j}.\label{eq: Ricci - Scalar Field}
\end{equation}

\subsection{Static Spherically Symmetric Solution}

Generalizing, one may take the trace of the stress tensor of matter
as the source of the scalar field. For a static point source, the
covariant d'Alambert equation Eq. (\ref{eq: Covariant d'Alambert Eq.})
with the stress tensor trace for a perfect fluid Eq. (\ref{eq: Trace Fluid})
has the solution

\begin{equation}
\phi=-\frac{GM}{r},\quad r>0\label{eq: Grav. Pot.}
\end{equation}
which were simplified by introducing the conformal factor Eq. (\ref{eq: Conformal Factor})
\begin{equation}
\sqrt{-g}=e^{-\nicefrac{2\phi}{c_{0}^{2}}}.
\end{equation}
The scalar field stress tensor is Eq. (\ref{eq: Stress Tensor Scl Field})

\begin{equation}
?T^{i}{}_{j}?^{\left(\phi\right)}=-\frac{1}{4\pi G}\left(\phi^{i}\phi_{j}-\frac{1}{2}\delta_{j}^{i}\phi^{r}\phi_{r}\right)=\frac{GM^{2}}{8\pi}\left[\begin{array}{cccc}
\frac{e^{\nicefrac{2\phi}{c_{0}^{2}}}}{r^{4}} & 0 & 0 & 0\\
0 & -\frac{e^{\nicefrac{2\phi}{c_{0}^{2}}}}{r^{4}} & 0 & 0\\
0 & 0 & \frac{e^{\nicefrac{2\phi}{c_{0}^{2}}}}{r^{4}} & 0\\
0 & 0 & 0 & \frac{e^{\nicefrac{2\phi}{c_{0}^{2}}}}{r^{4}}
\end{array}\right],\qquad T^{\left(\phi\right)}=\frac{GM^{2}e^{\nicefrac{2\phi}{c_{0}^{2}}}}{4\pi r^{4}}.\label{eq: Stress Tensor}
\end{equation}

From the Christoffel symbols $?\Gamma^{i}{}_{jk}?$ from tensor calculus
\begin{align}
R_{ij} & =\partial_{r}?\Gamma^{r}{}_{ij}?-\partial_{j}?\Gamma^{r}{}_{ir}?+?\Gamma^{r}{}_{ij}??\Gamma^{s}{}_{rs}?-?\Gamma^{s}{}_{ir}??\Gamma^{r}{}_{js}?,\nonumber \\
?\Gamma^{i}{}_{jk}? & =\frac{1}{2}g^{ir}\left(\partial_{k}g_{rj}+\partial_{j}g_{rk}-\partial_{r}g_{jk}\right)\label{eq: Christoffel Symbols}
\end{align}
and Einsteins field equations
\begin{equation}
?T^{i}{}_{j}?^{\left(\phi\right)}=\frac{1}{\kappa}\left(?R^{i}{}_{j}?-\frac{1}{2}\delta_{j}^{i}R\right)
\end{equation}
one can solve for the metric, which leads to \cite{Yilmaz}

\begin{equation}
ds^{2}=-e^{\nicefrac{2\phi}{c_{0}^{2}}}c_{0}^{2}dt^{2}+e^{-\nicefrac{2\phi}{c_{0}^{2}}}\left(dr^{2}+r^{2}d\theta^{2}+r^{2}\sin^{2}\left(\theta\right)d\varphi^{2}\right)\label{eq: Yilmaz Metric}
\end{equation}

It is noteworthy that it is free of additional singularities. Also,
for a classical field theory (e.g. gravitation, elasticity, electromagnetism,
fluid dynamics, and thermo dynamics), invariance of the potential
$\phi$ under a group (of transformations) is described by
\begin{equation}
\phi\rightarrow\phi+\alpha
\end{equation}
where $\alpha$ is a constant. Generalizing, for a function $f\left(\phi\right)$
to be invariant at two different locations $i$ and $j$ requires
that \cite{Rastall}
\begin{equation}
\frac{f\left(\phi_{i}\right)}{f\left(\phi_{j}\right)}=\frac{f\left(\phi_{i}+\alpha\right)}{f\left(\phi_{j}+\alpha\right)}
\end{equation}
which leads to
\begin{equation}
d\ln\left(f\left(\phi_{i}\right)\right)=d\ln\left(f\left(\phi_{j}\right)\right)=\mathrm{cst}
\end{equation}
with the solution
\begin{equation}
f\left(\phi_{i}\right)=f\left(\phi_{j}\right)e^{a\left(\phi_{i}-\phi_{j}\right)}
\end{equation}
where $a$ is a constant. Clearly, e.g. the Schwarzschild metric for
GR Eq. (\ref{eq: Schwarzschild Metric}) does not have this property.
Hence the introduction of ``test particles'' (in an ``external
field'') in GR text books; vs. interacting particles. Alternatively,
like Einstein concluded in 1945 \cite{Meaning_of_Rel.}, one may view
it as a leading order theory.

The stress tensor density in mixed form has a particularly simple
expression
\begin{equation}
?{\mathscr{T}}^{i}{}_{j}?^{\left(\phi\right)}=\frac{\sqrt{-g}}{\kappa}\left(?R^{i}{}_{j}?-\frac{1}{2}\delta_{j}^{i}R\right)=\frac{GM^{2}}{8\pi}\left[\begin{array}{cccc}
\frac{1}{r^{4}} & 0 & 0 & 0\\
0 & -\frac{1}{r^{4}} & 0 & 0\\
0 & 0 & \frac{1}{r^{4}} & 0\\
0 & 0 & 0 & \frac{1}{r^{4}}
\end{array}\right],\qquad\mathscr{T}^{\left(\phi\right)}=\frac{GM^{2}}{4\pi r^{4}}\label{eq: Stress Tensor Point Chg M}
\end{equation}
i.e., like a static point charge in electromagnetism.

Comparing with the Schwarzschild metric \cite{Schwarzschild}
\begin{equation}
?T^{i}{}_{j}?^{\left(\mathrm{m}\right)}=\left[\begin{array}{cccc}
0 & 0 & 0 & 0\\
0 & 0 & 0 & 0\\
0 & 0 & 0 & 0\\
0 & 0 & 0 & 0
\end{array}\right],\qquad T=0,\qquad r>0
\end{equation}
because the GR stress tensor only contains the sources for the gravitational
field; not the energy-momentum for the field itself.

Not surprisingly, much has been written on this topic since the inception
of the theory; e.g. the Levi-Civita - Einstein controversy 1917 \cite{Levi-Civita_Stress_Tensor}.
For a perspective, one may compare with electromagnetism, for which
the Maxwell stress and electro-magnetic field tensors are given by
\begin{equation}
?{\mathscr{T}}^{i}{}_{j}?^{\left(\mathrm{em}\right)}=\frac{1}{\mu_{0}}\left(\mathcal{F}^{ir}F_{jr}-\frac{1}{4}\delta_{j}^{i}\mathcal{F}^{rs}F_{rs}\right),\qquad F_{ij}=\partial_{i}A_{j}-\partial_{j}A_{i}
\end{equation}
and
\begin{equation}
\partial_{r}\mathscr{T}^{ir\left(\mathrm{em}\right)}=\mu_{0}\mathscr{J}^{i}.
\end{equation}
For e.g. a point charge $Q$ the stress tensor is

\begin{equation}
?T^{i}{}_{j}?^{\left(\mathrm{em}\right)}=\frac{Q^{2}}{32\pi^{2}\epsilon_{0}}\begin{bmatrix}-\frac{1}{r^{4}} & 0 & 0 & 0\\
0 & -\frac{1}{r^{4}} & 0 & 0\\
0 & 0 & \frac{1}{r^{4}} & 0\\
0 & 0 & 0 & \frac{1}{r^{4}}
\end{bmatrix},\qquad T^{\left(\mathrm{em}\right)}=0\label{eq: Stress Tensor Point Chg Q}
\end{equation}
which may be compared with the previous result Eq. (\ref{eq: Stress Tensor Point Chg M}).

\subsection{Static Spherically Symmetric Body with Charge: Reissner-Nordström
Metric}

Because the Maxwell stress tensor has zero trace, the electro-magnetic
field is not a source of the scalar field. Hence, it does not affect
the metric. However, it can contribute to the rest mass of the body;
so called Maxwell stresses.

For GR, Einstein's field equations Eq. (\ref{eq: Einstein's Field Eqs.})
for a charged spherically symmetric body with mass $M$ and charge
$Q$ are
\begin{equation}
R_{ij}-\frac{1}{2}g_{ij}R=\kappa T_{ij}^{\left(\mathrm{em}\right)}=\kappa\frac{Q^{2}}{32\pi^{2}\epsilon_{0}}\begin{bmatrix}-\frac{1}{r^{4}} & 0 & 0 & 0\\
0 & -\frac{1}{r^{4}} & 0 & 0\\
0 & 0 & \frac{1}{r^{4}} & 0\\
0 & 0 & 0 & \frac{1}{r^{4}}
\end{bmatrix}
\end{equation}
where the R.H.S is the Maxwell stress tensor for a point charge Eq.
(\ref{eq: Stress Tensor Point Chg Q}). The solution is \cite{Reissner,Nordstr=0000F6m_Metric}
\begin{equation}
ds^{2}=-\left(1-\frac{r_{\mathrm{S}}}{r}+\frac{r_{\mathrm{Q}}^{2}}{r^{2}}\right)c_{0}^{2}dt^{2}+\frac{1}{1-\frac{r_{\mathrm{S}}}{r}+\frac{r_{\mathrm{Q}}^{2}}{r^{2}}}dr^{2}+r^{2}d\theta^{2}+r^{2}\sin^{2}\left(\theta\right)d\varphi^{2}
\end{equation}
where
\begin{equation}
r_{\mathrm{S}}\equiv\frac{2GM}{c_{0}^{2}},\qquad r_{\mathrm{Q}}^{2}\equiv\frac{GQ^{2}}{4\pi\epsilon_{0}c_{0}^{4}}
\end{equation}
which has additional singularities for
\begin{equation}
r=\frac{1}{2}\left(r_{\mathrm{S}}\pm\sqrt{r_{\mathrm{S}}^{2}-4r_{\mathrm{Q}}^{2}}\right).
\end{equation}

\subsection{Lagrangian Formulation and Post-Newtonian Approximation}

A Lagrangian formulation is obtained from the Lagrangian

\begin{equation}
S=\int Ldt=-\int\frac{m_{0}c_{0}^{2}}{\gamma}dt=-\int m_{0}c_{0}^{2}\frac{ds}{c_{0}dt}dt\label{eq: SR Lagrangian}
\end{equation}
since \cite{Tonti}
\begin{equation}
L\equiv\int\bar{p}\left(\bar{v}\right)\cdot d\bar{v}=\int m_{0}\gamma vdv=-\frac{m_{0}c_{0}^{2}}{\gamma}
\end{equation}
and
\begin{equation}
\frac{ds}{c_{0}dt}=\frac{1}{\gamma}.
\end{equation}
For the Schwarzschild metric in isotropic form \cite{Eddington}
\begin{equation}
ds^{2}=-\frac{\left(1+\frac{\phi}{2c_{0}^{2}}\right)^{2}}{\left(1-\frac{\phi}{2c_{0}^{2}}\right)^{2}}c_{0}^{2}dt^{2}+\left(1-\frac{\phi}{2c_{0}^{2}}\right)^{4}\left(dr^{2}+r^{2}d\theta^{2}+r^{2}\sin^{2}\left(\theta\right)d\varphi^{2}\right)\label{eq: Schwarzschild}
\end{equation}
one obtains the Post-Newtonian approximation
\begin{equation}
L=-m_{0}c_{0}^{2}\frac{ds}{c_{0}dt}=-m_{0}c_{0}^{2}\sqrt{\frac{\left(1+\frac{\phi}{2c_{0}^{2}}\right)^{2}}{\left(1-\frac{\phi}{2c_{0}^{2}}\right)^{2}}-\left(1-\frac{\phi}{2c_{0}^{2}}\right)^{4}\frac{v^{2}}{c_{0}^{2}}}=-m_{0}\left(\frac{v^{2}}{2}-\phi-\frac{\phi^{2}}{2c_{0}^{2}}-\frac{3\phi v^{2}}{2c_{0}^{2}}+\frac{v^{4}}{8c_{0}^{2}}+\ldots\right)\label{eq: Lagrangian}
\end{equation}
where $\bar{v}\equiv\frac{d\bar{x}}{dt}$ and by ignoring the constant
term. The equations of motion are
\begin{equation}
\frac{d}{dt}\left(\partial_{v^{i}}L\right)-\partial_{i}L=m_{0}\left[\left(1-\frac{3\phi}{c_{0}^{2}}\right)\frac{d\bar{v}}{dt}+\left(1+\frac{\phi}{c_{0}^{2}}+\frac{3v^{2}}{2c_{0}^{2}}\right)\nabla\phi+\ldots\right]=0
\end{equation}
which simplify to
\begin{equation}
\frac{d\bar{v}}{dt}=-\left(1+\frac{4\phi}{c_{0}^{2}}\right)\nabla\phi+\ldots
\end{equation}
Similarly, for the exponential metric Eq. (\ref{eq: Yilmaz Metric})
 one obtains
\begin{equation}
L=-m_{0}c_{0}^{2}e^{\nicefrac{\phi}{c_{0}^{2}}}\sqrt{1-\left(\frac{v}{c_{0}e^{\nicefrac{2\phi}{c_{0}^{2}}}}\right)^{2}}=-m_{0}\left(\frac{v^{2}}{2}-\phi-\frac{\phi^{2}}{2c_{0}^{2}}-\frac{3\phi v^{2}}{2c_{0}^{2}}+\frac{v^{4}}{8c_{0}^{2}}+\ldots\right)\label{eq: SR -> GR 1}
\end{equation}
which agrees with Eq. (\ref{eq: Lagrangian}) to leading order in
the gravitational constant. Hence, by introducing this Lagrangian
to analytic mechanics, what is experimentally known for GR can be
accounted for.

\subsection{Generalized Nordström Theory}

In 1913 G. Nordström developed a self-consistent scalar field theory
for gravity with Minkowski metric $\eta_{ij}$ \cite{Nordstr=0000F6m}.
To quote A. Pais \cite{Pais}:
\begin{quote}
\textit{\textquotedblleft{}Though it was not to survive, it deserves
to be remembered as the first logically consistent relativistic field
theory of gravitation ever formulated.\textquotedblright{}}
\end{quote}
His field equations can be generated by the Lagrangian density \cite{Nordstr=0000F6m,Pais,Einstein_Nordstr=0000F6m}

\begin{equation}
\mathscr{L}=\frac{1}{8\pi G}\phi^{r}\phi_{r}-m_{0}c_{0}\left(1+\frac{\phi}{c_{0}^{2}}\right)\sqrt{-u_{r}u^{r}}=\frac{1}{8\pi G}\phi^{r}\phi_{r}-c_{0}\sqrt{-p_{r}p^{r}},\qquad p^{i}\equiv m_{0}\left(1+\frac{\phi}{c_{0}^{2}}\right)u^{i}.
\end{equation}
A variation of the field gives
\begin{equation}
\partial_{\phi}\mathscr{L}-\partial_{r}\left(\partial_{\phi_{r}}\mathscr{L}\right)=0
\end{equation}
which leads to the field equation

\begin{equation}
\square\phi=-4\pi Gm_{0}
\end{equation}
and the scalar field stress tensor

\begin{equation}
T^{ij\left(\phi\right)}=-\frac{1}{4\pi G}\left(\phi^{i}\phi^{j}-\frac{1}{2}\eta^{ij}\phi^{r}\phi_{r}\right)=-\phi^{j}\partial_{\phi_{i}}\mathscr{L}+\eta^{ij}\mathscr{L}
\end{equation}
with the trace
\begin{equation}
T^{\left(\phi\right)}=-\frac{1}{4\pi G}\phi^{r}\phi_{r}.
\end{equation}
Similarly, the matter stress tensor is
\begin{equation}
T^{ij\left(m\right)}=-m_{0}\left(1+\frac{\phi}{c_{0}^{2}}\right)u^{i}u^{j}
\end{equation}
with the trace
\begin{equation}
T^{\left(\mathrm{m}\right)}=m_{0}c_{0}^{2}\left(1+\frac{\phi}{c_{0}^{2}}\right).
\end{equation}
Hence, the field equations can be written
\begin{equation}
\square\phi=-\frac{4\pi G}{c_{0}^{2}}\frac{T^{\left(\mathrm{m}\right)}}{1+\frac{\phi}{c_{0}^{2}}}
\end{equation}
The divergence of the total stress tensor is zero (equations of motion)
\begin{equation}
\partial_{r}T^{rj\mathrm{\left(tot\right)}}=\partial_{r}\left(T^{rj\left(\phi\right)}+T^{rj\mathrm{\left(m\right)}}\right)=0.
\end{equation}
The field equations may be compared with Eq. (\ref{eq: Covariant d'Alambert Eq.})
for the generalized theory (in curved space-time)
\begin{equation}
\square\left(\bm{{\phi}}\right)=-\frac{4\pi G}{c_{0}^{2}}\mathscr{T}{}^{\left(\mathrm{m}\right)}=-\frac{4\pi G}{c_{0}^{2}}\sqrt{-g}T{}^{\left(\mathrm{m}\right)}=-\frac{4\pi G}{c_{0}^{2}e^{\nicefrac{2\phi}{c_{0}^{2}}}}T{}^{\left(\mathrm{m}\right)}=-\frac{4\pi G}{c_{0}^{2}}\frac{T^{\left(\mathrm{m}\right)}}{1+\frac{2\phi}{c_{0}^{2}}+\ldots}.\label{eq: Field Eqs. Gen. Nordstr=0000F6m}
\end{equation}
generated by the Lagrangian density
\begin{equation}
\mathscr{L}=\sqrt{-g}\left(\frac{1}{8\pi G}\phi^{r}\phi_{r}-m_{0}c_{0}\sqrt{-u_{r}u^{r}}\right)=e^{-\nicefrac{2\phi}{c_{0}^{2}}}\left(\frac{1}{8\pi G}\phi^{r}\phi_{r}-m_{0}c_{0}\sqrt{-u_{r}u^{r}}\right)
\end{equation}
and the metric Eq. (\ref{eq: Yilmaz Metric})
\begin{equation}
ds^{2}=-e^{\nicefrac{2\phi}{c_{0}^{2}}}c_{0}^{2}dt^{2}+e^{-\nicefrac{2\phi}{c_{0}^{2}}}\left(dr^{2}+r^{2}d\theta^{2}+r^{2}\sin^{2}\left(\theta\right)d\varphi^{2}\right).
\end{equation}

\section{Metric: Constitutive Relations for the Space-Time Continuum}
\begin{quote}
\textit{``As far as the laws of mathematics refer to reality, they
are not certain;}

\textit{and as far as they are certain, they do not refer to reality.''}

A. Einstein (1921) \cite{Einstein_1921}.
\end{quote}

\subsection{Metric for the Space-Time Continuum}

In 1907 A. Einstein concluded that the finite transformation for time
dilatations must be translation invariant (section 18) \cite{Time_Dilatation})
\begin{equation}
\Delta t=\Delta t_{0}e^{\nicefrac{a\Delta x}{c_{0}^{2}}}
\end{equation}
where $a=\mathrm{const.}$ is the acceleration. But oddly, when generalizing
to GR, limited the considerations to the infinitesimal part, his Eq.
(30)
\begin{equation}
\Delta t=\Delta t_{0}\left(1+\frac{a\Delta x}{c_{0}^{2}}+\ldots\right),
\end{equation}
by using his Eq. (30a); although GR is a fundamentally nonlinear theory.
Similarly, in 1911 he obtained the infinitesimal transformation for
the redshift due to a gravitational field \cite{Redshift}
\begin{equation}
f=f_{0}\left(1+\frac{\phi}{c_{0}^{2}}+\ldots\right).
\end{equation}
The finite transformation is \cite{Yilmaz}
\begin{equation}
f=f_{0}e^{\nicefrac{\phi}{c_{0}^{2}}}=f_{0}\left(1+\frac{\phi}{c_{0}^{2}}+\ldots\right).\label{eq: Redshift}
\end{equation}

Similarly, the rest-mass is not conserved in a gravitational field;
only mass ratios. Starting from the SR relation
\begin{equation}
E=m_{0}c_{0}^{2},
\end{equation}
Nordström obtained for an accelerated frame \cite{Nordstr=0000F6m}
\begin{equation}
mc_{0}^{2}=m_{0}c_{0}^{2}e^{\nicefrac{\phi}{c_{0}^{2}}}.
\end{equation}
Alternatively, the same result is obtained for the redshift Eq. (\ref{eq: Redshift})
of a de Brogli (``matter'') wave \cite{de-Broglie}
\begin{equation}
E=\mathrm{h}\upsilon,
\end{equation}
where $\mathrm{h}$ is Planck's constant, which leads to
\begin{equation}
E=mc_{0}^{2}=\mathrm{h}\upsilon=\mathrm{h}\upsilon_{0}e^{\nicefrac{\phi}{c_{0}^{2}}}=m_{0}c_{0}^{2}e^{\nicefrac{\phi}{c_{0}^{2}}}.\label{eq: Rest Mass Change}
\end{equation}

Also, in the 1911 paper he concluded that
\begin{equation}
c\left(\phi\right)=c_{0}\left(1+\frac{\phi}{c_{0}^{2}}\right).\label{eq: Speed of Light Einstein}
\end{equation}
Actually, since for light $ds=0$, the Schwarzschield metric in isotrophic
form Eq. (\ref{eq: Schwarzschild}) gives for the local speed of light
(section 83 (c)) \cite{Tolman} 
\begin{equation}
c\left(\phi\right)=\frac{dr}{dt}=\frac{1+\frac{\phi}{2c_{0}^{2}}}{\left(1-\frac{\phi}{2c_{0}^{2}}\right)^{3}}c_{0}=c_{0}\left(1+\frac{2\phi}{c_{0}^{2}}+\ldots\right)\label{eq: Speed of Light Schwarzschild}
\end{equation}
a factor 2 difference; because there is a length contraction as well
as a time dilatation (observed by ``radar echo delay'' experiments
\cite{Radar_Echo_Delay}). Similarly, for the exponential metric one
obtains
\begin{equation}
c\left(\phi\right)=\frac{dr}{dt}=e^{\nicefrac{2\phi}{c_{0}^{2}}}c_{0}=c_{0}\left(1+\frac{2\phi}{c_{0}^{2}}+\ldots\right)\label{eq: Local Speed of Light Exp}
\end{equation}
i.e., they agree to leading order. The exponential metric can be written
in the form
\begin{align}
ds^{2} & =-e^{\nicefrac{2\phi}{c_{0}^{2}}}c_{0}^{2}dt^{2}+e^{-\nicefrac{2\phi}{c_{0}^{2}}}\left(dr^{2}+r^{2}d\theta^{2}+r^{2}\sin^{2}\left(\theta\right)d\varphi^{2}\right)\nonumber \\
 & =-e^{-\nicefrac{2\phi}{c_{0}^{2}}}\left(c^{2}\left(\phi\right)dt^{2}-dr^{2}-r^{2}d\theta^{2}-r^{2}\sin^{2}\left(\theta\right)d\varphi^{2}\right)\label{eq: Conf. Metric}
\end{align}
for which its conformal structure becomes transparent.

\subsection{Light Propagation: Fermat's Principle}

The Lagrangian for light propagation is (Fermat's principle)
\begin{equation}
S=\int c_{0}dt=\int\frac{c_{0}}{c}ds=\int nds
\end{equation}
where $n$ is the refractive index
\begin{equation}
n\equiv\frac{c_{0}}{c}.
\end{equation}
To evaluate the bending of light, Einstein in his 1916 paper \cite{Einstein_GR}
used the local speed of light obtained for $ds=0$ 
\begin{equation}
c\left(\phi\right)=\frac{dr}{dt}=c_{0}\left(1-\frac{2GM}{c_{0}^{2}r}+\ldots\right)
\end{equation}
where we have used the Schwarzschield metric in isotrophic form Eq.
(\ref{eq: Schwarzschild}). If the light moves along the $x$-axis
in the $z=0$ plane the transverse differential deflection is
\begin{equation}
\frac{d\theta}{dx}=\frac{1}{c\left(\phi\right)}\frac{dc\left(\phi\right)}{dy}
\end{equation}
which gives
\begin{equation}
\frac{d\theta}{dx}\approx\frac{1}{c_{0}}\frac{dc\left(\phi\right)}{dy}=\frac{2GM}{c_{0}^{2}}\frac{R+y}{\left[x^{2}+\left(R+y\right)^{2}\right]^{\nicefrac{3}{2}}}\approx\frac{2GM}{c_{0}^{2}}\frac{R}{\left[x^{2}+R^{2}\right]^{\nicefrac{3}{2}}}
\end{equation}
where $R$ is the distance of closest approach. An integration over
$x$ gives
\begin{equation}
\theta=\intop_{-\infty}^{\infty}\frac{d\theta}{dx}dx=\frac{2GM}{c_{0}^{2}}\left[\frac{x}{R\sqrt{x^{2}+R^{2}}}\right]_{-\infty}^{\infty}=\frac{4GM}{c_{0}^{2}R}
\end{equation}
 as expected.

\subsection{Phenomenological Approach: Polarizable Vacuum Interpretation}

The Lagrangian density for the electromagnetic field is

\begin{equation}
\mathscr{L}_{\mathrm{em}}=-\frac{1}{4\mu_{0}}\mathscr{F}^{rs}F_{rs}
\end{equation}
where (constitutive relation for vacuum)
\begin{equation}
\mathscr{F}^{ij}=\frac{\sqrt{-g}}{\mu_{0}}g^{ir}g^{is}F_{rs},\qquad c_{0}=\frac{1}{\sqrt{\epsilon_{0}\mu_{0}}}.\label{eq: Const. Rel. e-m}
\end{equation}
From the relations
\begin{equation}
\bar{D}=\epsilon_{0}\bar{E},\qquad\bar{H}=\frac{1}{\mu_{0}}\bar{B}
\end{equation}
for the conformal metric Eq. (\ref{eq: Conf. Metric}) one may introduce
(vacuum polarization)
\begin{equation}
\varepsilon_{0}\rightarrow\varepsilon_{0}e^{-\nicefrac{2\phi}{c_{0}^{2}}},\qquad\mu_{0}\rightarrow\mu_{0}e^{-\nicefrac{2\phi}{c_{0}^{2}}}\label{eq: Vacuum Pol. 1}
\end{equation}
and it follows that
\begin{equation}
c_{0}=\frac{1}{\sqrt{\mu_{0}\varepsilon_{0}}}\rightarrow c\left(\phi\right)=c_{0}e^{\nicefrac{2\phi}{c_{0}^{2}}}.\label{eq: Vacuum Pol. 2}
\end{equation}
Landau-Lifshitz made a similar analysis (section 90) \cite{Landau-Lifshitz}.
However, because the latter is based on the approximate relations
\begin{equation}
F^{ij}=\frac{1}{\mu_{0}}g^{ir}g^{is}F_{rs},\qquad D^{\alpha}=\sqrt{-g_{00}}F^{0\alpha},\qquad H^{\alpha\beta}=-\sqrt{-g_{00}}F^{\alpha\beta},\qquad H_{\alpha}=-\frac{1}{2}\sqrt{\frac{g}{g_{00}}}\epsilon_{\alpha\mu\nu}H^{\mu\nu}
\end{equation}
vs. the constitutive relation Eq. (\ref{eq: Const. Rel. e-m}), it
leads to
\begin{equation}
\varepsilon_{0}\rightarrow\frac{\varepsilon_{0}}{\sqrt{-g_{00}}}=\varepsilon_{0}e^{-\nicefrac{\phi}{c_{0}^{2}}},\qquad\mu_{0}\rightarrow\frac{\mu_{0}}{\sqrt{-g_{00}}}=\mu_{0}e^{-\nicefrac{\phi}{c_{0}^{2}}},\qquad c_{0}\rightarrow c\left(\phi\right)=c_{0}e^{\nicefrac{\phi}{c_{0}^{2}}}=c_{0}\left(1+\frac{\phi}{c_{0}^{2}}+\ldots\right)
\end{equation}
vs. Eq. (\ref{eq: Speed of Light Schwarzschild}) and (\ref{eq: Local Speed of Light Exp}).

From the redshift Eq. (\ref{eq: Rest Mass Change}) it follows that
\begin{equation}
m_{0}c_{0}^{2}\rightarrow m\left(\phi\right)c^{2}\left(\phi\right)=\frac{m_{0}c_{0}^{2}}{\sqrt[4]{-g}}=m_{0}c_{0}^{2}e^{\nicefrac{\phi}{c_{0}^{2}}}\label{eq: Vacuum Pol. 3}
\end{equation}
but mass ratios are preserved. Similarly, the Compton wavelength is
reduced by
\begin{equation}
\lambda_{\mathrm{c}}\equiv\frac{\mathrm{h}}{m_{0}c_{0}}\rightarrow\frac{\lambda_{\mathrm{c}}}{\sqrt[4]{-g}}=\lambda_{\mathrm{c}}e^{\nicefrac{\phi}{c_{0}^{2}}}.
\end{equation}
However, the ``vacuum impedance'' remains invariant
\begin{equation}
Z_{0}\equiv\sqrt{\frac{\mu_{0}}{\varepsilon_{0}}}\rightarrow\sqrt{\frac{\mu_{0}}{\varepsilon_{0}}}
\end{equation}
so the fine structure constant does not change
\begin{equation}
\alpha\equiv\frac{e_{0}^{2}}{4\pi\varepsilon_{0}c_{0}\hbar}=\frac{e_{0}^{2}Z_{0}}{4\pi\hbar}.
\end{equation}

Collecting the results, from the SR Lagrangian Eq. (\ref{eq: SR Lagrangian})
\begin{equation}
S=\int Ldt=-\int m_{0}c_{0}^{2}\frac{d\tau}{dt}dt=-\int\frac{m_{0}c_{0}^{2}}{\gamma}dt.
\end{equation}
the generalized Lagrangian is simply
\begin{align}
L=-\frac{m_{0}c_{0}^{2}}{\gamma}\rightarrow-m\left(\phi\right)c^{2}\left(\phi\right)\sqrt{1-\frac{v^{2}}{c^{2}\left(\phi\right)}} & =-m_{0}c_{0}^{2}e^{\nicefrac{\phi}{c_{0}^{2}}}\sqrt{1-\left(\frac{v}{c_{0}e^{\nicefrac{2\phi}{c_{0}^{2}}}}\right)^{2}}\nonumber \\
 & =-m_{0}c_{0}^{2}\sqrt{e^{\nicefrac{2\phi}{c_{0}^{2}}}-e^{-\nicefrac{2\phi}{c_{0}^{2}}}\left(\frac{v}{c_{0}}\right)^{2}}\label{eq: SR -> GR 2}
\end{align}
in agreement with the earlier obtained Eq. (\ref{eq: SR -> GR 1}).

The equations of motion are
\begin{equation}
\frac{d}{dt}\left(\partial_{v^{i}}L\right)-\partial_{i}L=\frac{d\bar{p}}{dt}+m_{0}\left[1+\left(\frac{v}{c_{0}e^{\nicefrac{2\phi}{c_{0}^{2}}}}\right)^{2}\right]e^{\nicefrac{2\phi}{c_{0}^{2}}}\gamma_{\phi}\nabla\phi=0\label{eq: Eqs. of Motion PV}
\end{equation}
where we have introduced
\begin{equation}
\bar{p}\equiv m_{0}\gamma_{\phi}\bar{v}e^{\nicefrac{-2\phi}{c_{0}^{2}}},\qquad\gamma_{\phi}\equiv\frac{1}{e^{\nicefrac{\phi}{c_{0}^{2}}}\sqrt{1-\left(\frac{v}{c_{0}e^{\nicefrac{2\phi}{c_{0}^{2}}}}\right)^{2}}}.
\end{equation}

To summarize:
\begin{itemize}
\item The gravitational effects on Maxwell\textquoteright{}s equations can
be interpreted as a \textquotedblleft{}vacuum polarization\textquotedblright{}
by the gravitational field governed by Eqs. (\ref{eq: Vacuum Pol. 1})
and (\ref{eq: Vacuum Pol. 2}).
\item The theory is conformally invariant \cite{Dicke_1} but not scale
invariant (a dilatation symmetry); because the relativistic fluid
stress tensor has non-zero trace Eq. (\ref{eq: Trace Fluid}). Hence,
e.g. the ``vacuum impedance'' and fine structure constant are unaffected
whereas the Compton wavelength is changed; by the local gravitational
field.
\item The generalized Lagrangian can be obtained by: either by introducing
the conformal metric Eq. (\ref{eq: Conf. Metric}) or by retaining
the Minkowski (flat) space-time but with a varying the local speed
of light and rest masses governed by Eqs. (\ref{eq: Vacuum Pol. 2})
and (\ref{eq: Vacuum Pol. 3}); leading to the same Lagrangian Eqs.
(\ref{eq: SR -> GR 1}) and (\ref{eq: SR -> GR 2}). A matter of perspective.
\item GR can be accounted for in analytic mechanics by introducing the generalized
Lagrangian.
\end{itemize}
The latter approach dates back to H. Wilson 1921 and R. Dicke 1957
\cite{Wilson,Dicke_2}. One may compare with e.g. fluid dynamics:
Euler vs. Lagrange (moving frame) formulation. Or the theory of elasticity:
the stress tensor in the original coordinates before vs. after deformation;
due to the introduction of strain, see e.g. the Cosserat brothers
1909 \cite{Cosserat}.

\section{Lagrangian Density, Field Equations, and Equations of Motion}

The Lagrangian density $\mathscr{L}$ for the scalar, matter, and
electromagnetic fields is
\begin{align*}
\mathscr{L} & =\sqrt{-g}\left(\frac{1}{8\pi G}g^{rs}\phi_{r}\phi_{s}-\frac{1}{2}\rho g^{rs}u_{r}u_{s}-\frac{1}{4\mu_{0}}g^{rs}g^{tu}F_{rt}F_{su}\right)\\
 & =\frac{1}{8\pi G}e^{-\nicefrac{2\phi}{c_{0}^{2}}}\phi^{r}\phi_{r}-\frac{1}{2}\rho e^{-\nicefrac{2\phi}{c_{0}^{2}}}u^{r}u_{r}-\frac{e^{-\nicefrac{2\phi}{c_{0}^{2}}}}{4\mu_{0}}F^{rs}F_{rs}
\end{align*}
and
\begin{equation}
p^{i}\equiv\sqrt{-g}\rho u^{i}=\rho e^{-\nicefrac{2\phi}{c_{0}^{2}}}u^{i}
\end{equation}
with the conformal metric Eq. (\ref{eq: Conf. Metric}) (the constitutive
relations for space-time continuum)

\begin{align}
ds^{2} & =-e^{\nicefrac{2\phi}{c_{0}^{2}}}c_{0}^{2}dt^{2}+e^{-\nicefrac{2\phi}{c_{0}^{2}}}\left(dr^{2}+r^{2}d\theta^{2}+r^{2}\sin^{2}\left(\theta\right)d\varphi^{2}\right)\nonumber \\
 & =-e^{-\nicefrac{2\phi}{c_{0}^{2}}}\left(c^{2}\left(\phi\right)dt^{2}-dr^{2}-r^{2}d\theta^{2}-r^{2}\sin^{2}\left(\theta\right)d\varphi^{2}\right)\nonumber \\
 & =-\left(e^{\nicefrac{2\phi}{c_{0}^{2}}}-\frac{v^{2}}{c_{0}^{2}}e^{-\nicefrac{2\phi}{c_{0}^{2}}}\right)c_{0}^{2}dt^{2}=-e^{\nicefrac{2\phi}{c_{0}^{2}}}\left(1-\frac{v^{2}}{c_{0}^{2}e^{\nicefrac{4\phi}{c_{0}^{2}}}}\right)c_{0}^{2}dt^{2}=-\frac{1}{\gamma_{\phi}^{2}}c_{0}^{2}dt^{2}
\end{align}
where
\begin{equation}
\gamma_{\phi}\equiv\frac{1}{e^{\nicefrac{\phi}{c_{0}^{2}}}\sqrt{1-\left(\frac{v}{c_{0}e^{\nicefrac{2\phi}{c_{0}^{2}}}}\right)^{2}}}
\end{equation}
The equations of motion Eq. (\ref{eq: Eqs. of Motion Fluid 2}) can
be written
\begin{align}
u^{r}\nabla_{r}\left(\sqrt{-g}u_{i}\right) & =\frac{d}{ds}\left(\sqrt{-g}u_{i}\right)-?\Gamma_{i}{}{}_{rs}?\left(\sqrt{-g}\right)^{2}u^{r}u^{s}+\frac{c_{0}^{2}\sqrt{-g}}{2}g^{ir}\partial_{r}\ln\left(\rho\right)\nonumber \\
 & =\frac{d}{ds}\left(\sqrt{-g}u_{i}\right)-\frac{\left(\sqrt{-g}\right)^{2}}{2}\left(\partial_{i}g_{rs}\right)u^{r}u^{s}+\frac{c_{0}^{2}\sqrt{-g}}{2}\partial_{i}\ln\left(\rho\right)=0.
\end{align}
For a particle with mass $m_{0}$ 
\begin{align*}
\frac{d}{ds}\left(\sqrt{-g}u_{i}\right)-\frac{\left(\sqrt{-g}\right)^{2}}{2}\left(\partial_{i}g_{rs}\right)u^{r}u^{s} & =\frac{1}{m_{0}}\frac{dp_{i}}{ds}+\frac{1}{2}\left(\partial_{i}g_{rs}\right)u^{r}u^{s}=\frac{\gamma_{\phi}}{m_{0}c_{0}^{2}}\frac{dp_{i}}{dt}-\frac{\gamma_{\phi}^{2}}{2c_{0}^{2}}\left(\partial_{i}g_{rs}\right)\frac{dx^{r}}{dt}\frac{dx^{s}}{dt}\\
 & =\frac{\gamma_{\phi}}{m_{0}c_{0}^{2}}\frac{dp_{i}}{dt}+\frac{\gamma_{\phi}^{2}e^{\nicefrac{2\phi}{c_{0}^{2}}}}{c_{0}^{2}}\left[1+\left(\frac{v}{c_{0}e^{\nicefrac{2\phi}{c_{0}^{2}}}}\right)^{2}\right]\partial_{i}\phi=0
\end{align*}
it follows that
\begin{equation}
\frac{dp_{i}}{dt}+m_{0}\left[1+\left(\frac{v}{c_{0}e^{\nicefrac{2\phi}{c_{0}^{2}}}}\right)^{2}\right]e^{\nicefrac{2\phi}{c_{0}^{2}}}\gamma_{\phi}\partial_{i}\phi=0
\end{equation}
i.e., the same result as obtained from the corresponding Lagrangian
Eq. (\ref{eq: SR -> GR 1}) or (\ref{eq: SR -> GR 2}) and the equations
of motion Eq. (\ref{eq: Eqs. of Motion PV}).

The field equations for the scalar field are
\begin{equation}
\square\boldsymbol{\mathbf{\phi}}=-\frac{4\pi G}{c_{0}^{2}}\phi^{r}\phi_{r}+4\pi G\rho e^{\nicefrac{-2\phi}{c_{0}^{2}}}+\frac{4\pi G}{c_{0}^{2}}\frac{\varepsilon_{0}e^{\nicefrac{-2\phi}{c_{0}^{2}}}}{2}F^{rs}F_{rs}.
\end{equation}
Since the trace for the stress tensors for the scalar and electromagnetic
fields are zero (massless), by a suitable choice of gauge
\begin{equation}
\phi\rightarrow\phi+f\left(x\right)
\end{equation}
and

\begin{equation}
F_{ij}\equiv\partial_{i}A_{j}-\partial_{j}A_{i},\qquad A_{i}\rightarrow A_{i}+\partial_{i}g\left(x\right)
\end{equation}
they simplify to
\begin{equation}
\square\mathbf{\boldsymbol{\phi}}=4\pi G\rho e^{\nicefrac{-2\phi}{c_{0}^{2}}}.
\end{equation}
Maxwell's equations are equivalent to a medium (space-time continuum)
with local speed of light $c\left(\phi\right)$.

\section{Conclusions}

\noindent By introducing a scalar field/potential, and relating it
algebraically to the Ricci tensor, using Einstein's field equations,
a metric with an exponential dependance on the potential is obtained;
which is free of additional singularities. Hence, although the theory
is fundamentally nonlinear, the scalar field/potential provides an
analytic framework for interacting particles described by linear superposition.
The stress tensor for the scalar field includes both the sources of,
i.e., matter, and the energy-momentum for the gravitational field.
The stress tensor has zero covariant and ordinary divergence. Hence,
the energy-momentum for the gravitational field and sources are conserved.
The theory's predictions agree with the experimental results for General
Relativity. By introducing the corresponding Lagrangian in analytic
mechanics, what is experimentally known for GR can be accounted for.
The theory is essentially a generalization of Nordström's theory \cite{Nordstr=0000F6m,Pais,Einstein_Nordstr=0000F6m};
replacing the Minkowski (flat) metric with the exponential (curved)
metric Eq. (\ref{eq: Yilmaz Metric}). Alternatively, the Minkowski
metric can be retained by introducing a ``vacuum polarization''
due to the gravitational field that leads to varying local speed of
light and rest masses. The theory is conformally invariant but not
scale invariant (a dilatation symmetry); e.g. the ``vacuum impedance''
and fine structure constant are unaffected but the Compton wavelength
is changed by the local field.

\end{document}